\documentclass[a4paper]{article}

\usepackage{amsmath}
\usepackage{amsfonts}
\usepackage{amssymb}
\usepackage[dvips]{graphicx}
\usepackage{psfrag}

\DeclareFontFamily{OT1}{rsfs}{}
\DeclareFontShape{OT1}{rsfs}{m}{n}{ <-7> rsfs5 <7-10> rsfs7 <10-> rsfs10}{}
\DeclareMathAlphabet{\mycal}{OT1}{rsfs}{m}{n}

\def\scri{{\mycal I}}
\newcommand{\field}[1]{\mathbb{#1}}
\def\Real{\field{R}}
\newcommand{\romand}{{\rm d}}
\def\Lspace{\text{L}}
\def\Hspace{\text{H}}
\def\exter{\text{ext}}
\def\inter{\text{int}}

\title{Hamiltonian description of radiation phenomena:\\
       Trautman-Bondi energy and corner conditions
       }
\author{Witold Chmielowiec and Jerzy Kijowski\\
        Center for Theoretical Physics, Polish Academy of Sciences,\\
        Al. Lotnik\'ow 32/46, 02-668 Warsaw, Poland\\
        (e-mails: wchmiel@cft.edu.pl, kijowski@cft.edu.pl)
        }
\date{}

\begin{document}

\maketitle

\begin{abstract}
    Cauchy initial value problem on a hyperboloid is
    proved  to define a Hamiltonian system,
    provided the radiation data at null
    infinity are also taken into account, as a part of Cauchy data.
    The ``Trautman-Bondi mass'', supplemented by the ``already radiated
    energy'' assigned to radiation data,
    plays role of the Hamiltonian function. This approach
    leads to correct description of the corner conditions.
\end{abstract}

\noindent {\bf Keywords:} Trautman-Bondi energy, wave equation,
initial-characteristic value problem, Hamiltonian field theory.

\section{Introduction}
\label{sec:Introduction}

The notion of energy in the radiation regime (in our paper
referred to as the Trautman-Bondi energy) has been introduced in
Einstein's theory of gravity by Trautman \cite{Tra} and
independently by Bondi \cite{Bon}. It measures that part of the
gravitational energy of an isolated system, which ``has not yet
been radiated''. In conformal spacetime compactification, the
Trautman-Bondi energy may be assigned to any space-like
hypersurface having a regular intersection with the conformal
boundary $\scri$ (null infinity, or the {\em scri}) \cite{ChJSL}. Due to
radiation, the Trautman-Bondi energy -- unlike the total (ADM)
energy --  is not conserved, but is {\em decreasing} because it
may be partially radiated in a form of gravitational waves.

As shown in \cite{ChJK, Jez}, the validity of the Trautman-Bondi
energy goes far beyond the gravitational context and may be used
in any hyperbolic field theory, also special-relativistic one. In
particular, it has a beautiful Hamiltonian interpretation. The
goal of the present paper is to apply this idea to the scalar
field theory, where the Cauchy data of the system are assigned to
a hyperboloid. Field evolution on hyperboloids is proved to be a
Hamiltonian system, if we complete Cauchy data by the {\em
radiation data} at the {\em scri}, and supplement the
Trautman-Bondi energy with the corresponding {\em radiation
energy}. The sum of the two defines the total Hamiltonian function
of the ``matter + radiation'' system. But to ``tailor'' the two
disjoint objects: 1) the field Cauchy data on a hyperboloid and 2)
the radiation data on the {\em scri}, into a single object,
appropriate compatibly conditions (often called ``corner
conditions'' \cite{ChL, ChSL}) must be imposed. We propose a
universal approach which solves all these issues.

\section{Hamiltonian description of Cauchy initial
value problem on a hyperboloid}
\label{sec:CauchyInitialValueProblemOnAHyperboloid}

Hamiltonian description of any field dynamics is based on a ``3 +
1'' foliation of spacetime. Leaves of the foliation are labeled by
a parameter called the ``time variable''. Phase space of this
dynamics is composed of all the possible field Cauchy data on a
given leaf. The ``3 + 1'' decomposition provides also an
identification between different leaves of the foliation, which
makes the field dynamics and its Hamiltonian function uniquely
defined. Here, we consider the Hamiltonian description within a
simple model: the massless scalar field, satisfying the wave
equation in two- or four-dimensional Minkowski spacetime. This
means that the space is one-dimensional: $x\in \Real$ or
three-dimensional: $x\in\Real^3$. Contrary to the standard
(``ADM'') formulation, the field initial data will not be assigned
to spatially flat Cauchy surfaces, but to hypersurfaces which
extend to null infinity, namely spacelike hyperboloids. Naively,
such a Cauchy problem cannot be described as a Hamiltonian system:
future evolution of the system is well defined, but the past
evolution is absolutely non-unique and may be arbitrarily modified
by radiation. Nevertheless, we are going to show, that the
Hamiltonian formulation of the field evolution is possible. For
this purpose we have to complete Cauchy data on a hyperboloid by
appropriate {\em radiation data} at light infinity.

For pedagogical reasons, we begin our analysis with a (much
simpler) finite case, where we restrict field dynamics to a finite
light-cone and describe initial data on its (characteristic!)
boundary and on the finite part of the hyperboloid, contained
within the cone. At the end, we may consider the limiting case (in
our notation, this corresponds to $\epsilon \to 0$), where the
finite cone is shifted to infinity. This way we obtain the
Hamiltonian description of initial data on the entire hyperboloid
and on the \textit{scri} $\scri^+$ (conformal boundary of the
spacetime). In the subsequent Section we describe the
two-dimensional ``toy model''. The complete, four-dimensional
theory is analyzed in Section \ref{sec:13case}.

\subsection{Two-dimensional Minkowski spacetime}
\label{sec:11case}

Let us consider a one-parameter family of past oriented light
cones in the two-dimensional Minkowski space time:
\begin{equation}\label{eq:cone}
  \mycal{ C}_\epsilon^-:=
  \left\{ (t,x):\; x \in {\mathbb R}\, ,\; \frac1\epsilon - t > |x| \right\},
\end{equation}
where $\frac1\epsilon>1$ is the time coordinate of the vertex of
$\mycal{ C}_\epsilon^-$. We introduce new coordinates $(\tau , \xi
)$ connected with Minkowski coordinates $(t , x)$ in the following
way:
\begin{align}
  \label{eq:t_xi}
  t &=  \frac1\epsilon + \left(\frac{1+\xi^2}{1-\xi^2}-\frac1\epsilon
  \right)e^{-\epsilon\tau},
  \\
  \label{eq:x_xi}
  x &= \frac{2\xi}{1-\xi^2} e^{-\epsilon\tau} \;,
\end{align}
where $\tau \in {\mathbb R}^1$. For $| \xi | \leq
\frac{1-\epsilon}{1+\epsilon}$ new coordinates parameterize the
whole cone $\mycal{ C}_\epsilon^-$.
\begin{figure}[htbp]
    \centering
    \psfrag{x}{$x$}
    \psfrag{t}{$t$}
    \psfrag{a}{$\frac{1}{\epsilon}$}
    \psfrag{t0}{$\tau=0$}
    \psfrag{t1}{$\tau=\text{const.}$}
    \psfrag{x0}{$\xi=-\frac{1-\epsilon}{1+\epsilon}$}
    \psfrag{x1}{$\xi=\text{const.}$}
    \psfrag{x2}{$\xi=\frac{1-\epsilon}{1+\epsilon}$}
    \includegraphics[width=0.6\textwidth]{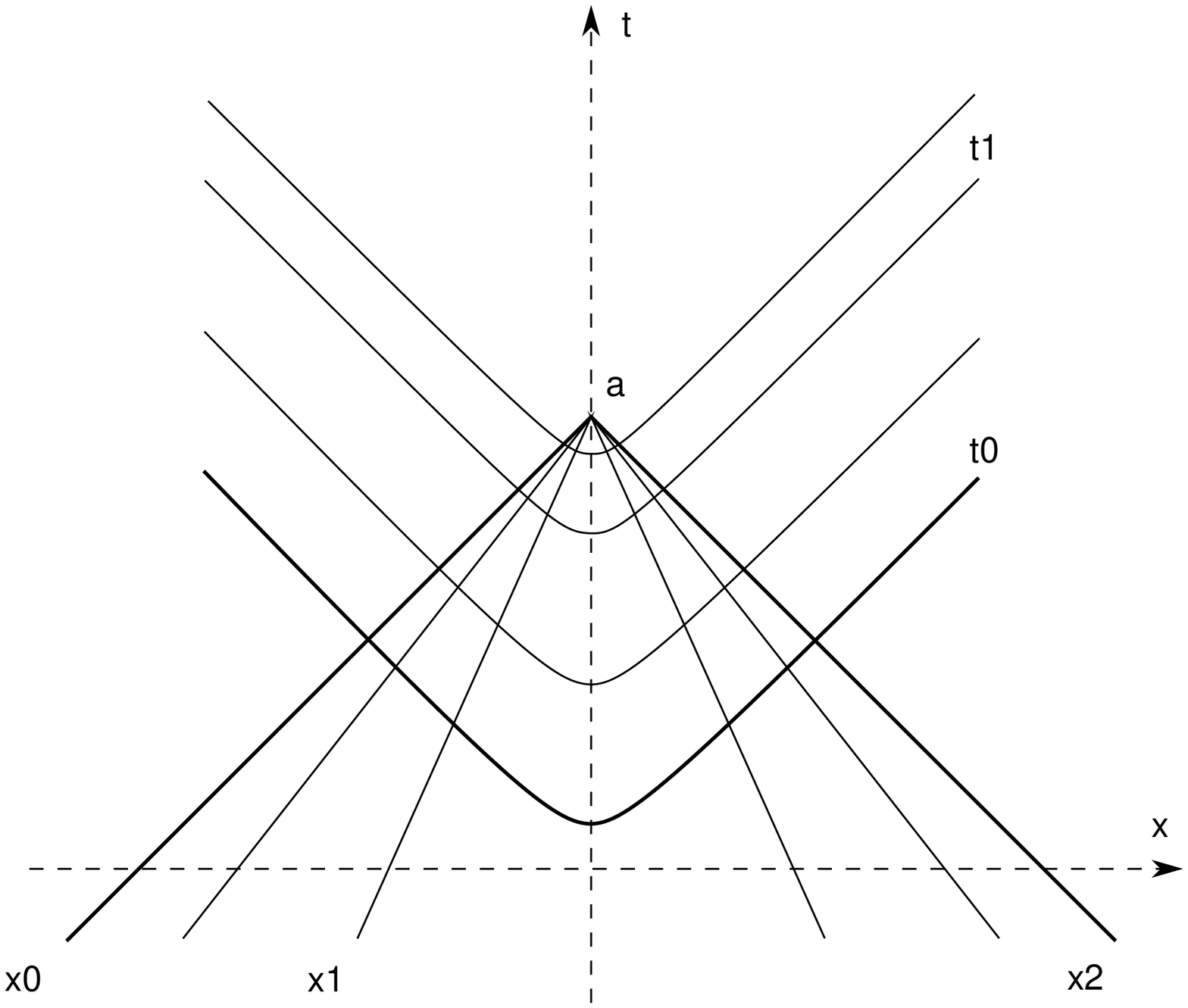}
    \label{fig:hyperbol}
\end{figure}
Moreover, surfaces $\{\tau= $ const.$\}$ correspond to
hyperboloids. In order to describe field dynamics in a Hamiltonian
way, we begin with the standard, relativistic Lagrangian for wave
equation:
\begin{equation}\label{eq:Lagr-st_we}
  L = - \frac 12 \sqrt{|\det g |}
  g^{\mu\nu} (\partial_\mu \varphi )(\partial_\nu \varphi ) =
  \frac 12 \left\{ \left( \partial_t \varphi \right)^2 -
  \left( \partial_x \varphi \right)^2 \right\} \;.
\end{equation}
Expressing the Lagrangian density ``$L \cdot \textrm{d}^2x$'' in
new coordinates we obtain:
\begin{equation}
   \label{eq:Lagr-st}
  L \cdot  \textrm{d}^2x = {\mathcal L} \cdot  \textrm{d}^2\xi \;,
\end{equation}
where
\begin{equation}
  \label{eq:Lagr-st-xi}
  {\mathcal L}  =
  \frac{\left( \frac{\partial \varphi}{\partial\tau} + \xi
  \frac{\partial \varphi}{\partial\xi}\right)^2}{1-\epsilon+(1+\epsilon)\xi^2}
  -\frac1 4 \left[{1-\epsilon+(1+\epsilon)\xi^2} \right]
  \left(\frac{\partial \varphi}{\partial\xi}\right)^2\, .
\end{equation}
Denoting
\begin{equation*}
  \kappa := \frac 1 2[ 1-\epsilon+(1+\epsilon)\xi^2]
\end{equation*}
we obtain the autonomous (i.e. $\tau$-independent) Lagrangian
function:
\begin{equation}
  \label{eq:Lagr-kappa}
  {\mathcal L}  =
  \frac{1}{2\kappa} (\partial_\tau \varphi + \xi\partial_\xi \varphi )^2
  -\frac1 2 \kappa (\partial_\xi\varphi)^2\, .
\end{equation}

Now, the standard procedure leading from the Lagrangian to the
Hamiltonian description may be applied. We first define conjugate
momenta:
\begin{equation}\label{eq:pi^mu}
  \pi^\mu := \frac {\partial \mathcal{L}}{\partial \varphi_\mu }\; ,
\end{equation}
where $\varphi_\mu:=\partial_\mu \varphi$, and calculate the
variation of the Lagrangian:
\begin{equation}\label{eq:delta-cL}
  \delta \mathcal{L} = \frac {\partial \mathcal{L}}{\partial \varphi} \delta
  \varphi + \pi^\mu \delta \varphi_\mu = \partial_\mu
  \left( \pi^\mu \delta \varphi
  \right) + \left( \frac {\partial \mathcal{L}}{\partial \varphi} -
  \partial_\mu \pi^\mu \right) \delta \varphi \;.
\end{equation}
Field equation (in our case it is always the wave equation
$\displaystyle \Box \varphi = 0$) is equivalent to vanishing of
the Euler-Lagrange term in \eqref{eq:delta-cL}, hence, equivalent
to the following equation
\begin{equation}\label{eq:delta-cL1}
  \delta \mathcal{L} =  \partial_\mu \left( \pi^\mu \delta \varphi
  \right)
   \, .
\end{equation}
Integrating \eqref{eq:delta-cL1} over the volume
$V_\epsilon:=\big\{ \xi :
|\xi|<\frac{1-\epsilon}{1+\epsilon}\big\}$ on the Cauchy surface
$\Sigma=\{ \tau =\text{const.} \}$, we obtain an identity valid for
fields satisfying wave equation
\begin{align*}
    \delta \int_{V_\epsilon} \mathcal{L} \romand\xi
    &=
    \int_{V_\epsilon}
    \left( \pi \delta \phi \right)^{\cdot}
    \romand\xi + \int_{\partial V_\epsilon} \left( \pi^1 \delta \phi \right)
    \romand\sigma_1\; ,
\end{align*}
where "dot" denotes derivative with respect to the new time
variable $\tau$, while $\phi$ is the restriction of the field
$\varphi$ to the Cauchy surface $\Sigma=\{\tau=\text{const.}\}$:
\begin{align}
  \notag
    \phi(\tau,\xi)&=\varphi\big(t(\tau,\xi), x(\tau,\xi)\big)\Big|_{\Sigma}
    \\
    &=\varphi\left(\tfrac1\epsilon
    + \big(\tfrac{1+\xi^2}{1-\xi^2}-\tfrac1\epsilon
    \big)e^{-\epsilon\tau}, \tfrac{2\xi}{1-\xi^2} e^{-\epsilon\tau}\right),
\end{align}
for $|\xi|\leq\frac{1-\epsilon}{1+\epsilon}$. The time component
of the momentum: $\pi:=\pi^0$, provides, together with $\phi$, the
complete description of Cauchy data on this surface. Legendre
transformation between $\dot{\phi}$ and $\pi$ gives us:
\begin{align}
  \label{eq:dH_int}
    -\delta H_{V_\epsilon}
    = \int_{V_\epsilon} ( \dot{\pi}\delta \phi
    -\dot{\phi}\delta \pi) \romand\xi +
    \big[\pi^1\delta\phi\big]_{\partial V_\epsilon},
\end{align}
where the Hamiltonian $H_{V_\epsilon}$ is defined by
\begin{align}
  \label{eq:H_int}
  H_{V_\epsilon}(\phi , \pi ) &:=  \int_{V_\epsilon}
  \big( \pi \dot{\phi} -   \mathcal{L} \big) \romand\xi \ .
\end{align}
Canonical structure in the space of Cauchy data is given by the
standard  symplectic form:
\begin{align}\label{eq:symp_phi_pi}
    \omega_{V_\epsilon} &:= \int_{V_\epsilon}
    (\delta\pi\wedge\delta\phi)\romand\xi.
\end{align}
Formulae \eqref{eq:Lagr-kappa} and \eqref{eq:pi^mu} imply the
following relations between ``velocity'' $\dot{\phi}$ and
momentum~$\pi$:
\begin{equation}
  \label{eq:pi^0}
  \pi := \pi^0 =  \frac {\partial \mathcal{L}}{\partial\dot\varphi} =
  \kappa^{-1} \left(\dot{\phi} + \xi
  \partial_\xi \phi\right) \, ,
\end{equation}
Taking into account \eqref{eq:Lagr-kappa} and \eqref{eq:pi^0}, the
Hamiltonian \eqref{eq:H_int} may be written explicitly in terms of
the Cauchy data on $V_\epsilon$
\begin{align}
  \label{eq:H_int(phi,pi)}
  H_{[-\frac{1-\epsilon}{1+\epsilon},\frac{1-\epsilon}{1+\epsilon}]}(\phi,\pi)
  &=
  \frac 12
  \int_{-\frac{1-\epsilon}{1+\epsilon}}^{\frac{1-\epsilon}{1+\epsilon}}
  \left\{ \kappa \left( \pi -\kappa^{-1}\xi
  \frac {\partial \phi}{\partial \xi}\right)^2 +
  \left(\kappa-\kappa^{-1}\xi^2\right)
  \left(\frac{\partial \phi}{\partial \xi}\right)^2
  \right\} \romand\xi \;.
\end{align}
The factor $\kappa-\kappa^{-1}\xi^2 $ is positive for
$\kappa-|\xi|>0$. Moreover, we have:
\begin{align}
    \kappa-|\xi|=\frac12
    (1+\epsilon)(1-|\xi|)\bigg(\frac{1-\epsilon}{1+\epsilon}-|\xi|\bigg) \ .
\end{align}
This implies positivity of the Hamiltonian
\eqref{eq:H_int(phi,pi)} for
$|\xi|<\frac{1-\epsilon}{1+\epsilon}$, i.e~inside the cone
$\mycal{ C}_\epsilon^-$. The following Hamiltonian equations
(equivalent to Euler-Lagrange equations) may be derived from the
Hamiltonian \eqref{eq:H_int(phi,pi)}:
\begin{align}
    \dot{\phi} &= \kappa \pi - \xi \partial_{\xi} \phi \ ,\\
    \dot{\pi} &= \partial_{\xi}(\kappa \partial_{\xi} \phi)
    - \partial_{\xi} (\xi\pi)
    \ ,
\end{align}
provided no boundary terms remain after the integration by part of
its variation, which, {\em a priori}, is not true! This apparent
paradox is implied by the fact, that evolution of Cauchy data on a
hyperboloid is well defined only forward in time and, whence, does
not correspond {\em a priori} to any Hamiltonian system. To
overcome this difficulty, we take into account missing data on the
light cone below hyperboloid and treat it as a part of Cauchy data
(cf.~\cite{ChJK}). For this purpose we extend parametrization
\eqref{eq:t_xi}, \eqref{eq:x_xi} beyond the volume
$V_\epsilon=\big\{\xi : |\xi| <
\frac{1-\epsilon}{1+\epsilon}\big\}$, taking into account also the
corresponding points on the boundary of the cone:
\begin{align}
  \label{eq:t=-x}
    t = \frac1\epsilon - x &
     := \frac1\epsilon - \frac12\Big(\frac1\epsilon - \epsilon\Big)
     e^{-\epsilon\tau + \epsilon\xi - \frac{\epsilon(1-\epsilon)}{1+\epsilon}}
     \quad
     \text{for}\ \xi\geq\frac{1-\epsilon}{1+\epsilon} \;,
    \\
    \label{eq:t=x}
    t = \frac1\epsilon + x &
     := \frac1\epsilon -  \frac12\Big(\frac1\epsilon - \epsilon\Big)
     e^{-\epsilon\tau - \epsilon\xi - \frac{\epsilon(1-\epsilon)}{1+\epsilon}}
     \quad
     \text{for}\ \xi \leq -\frac{1-\epsilon}{1+\epsilon} \;,
\end{align}
and consider the data $(\phi,\pi)$ on the entire surface $\Sigma =
\{\tau=\text{const.},\xi\in \Real\}$. Thus
\begin{align}
  \notag
    \phi(\tau,\xi)&=\varphi\big(t(\tau,\xi), x(\tau,\xi)\big)
    \Big|_{\partial{\mycal C}_\epsilon^{-}}
    \\
    \label{eq:phi_R_cone}
    &=\varphi\Big(\tfrac1\epsilon - \tfrac12\big(\tfrac1\epsilon
    - \epsilon\big)
     e^{-\epsilon\tau + \epsilon\xi - \frac{\epsilon(1-\epsilon)}{1+\epsilon}},
     \tfrac12\big(\tfrac1\epsilon - \epsilon\big)
     e^{-\epsilon\tau + \epsilon\xi
     - \frac{\epsilon(1-\epsilon)}{1+\epsilon}}\Big)
\end{align}
for $\xi    \geq    \frac{1-\epsilon}{1+\epsilon}$,
\begin{align}
  \notag
    \phi(\tau,\xi)&=\varphi\big(t(\tau,\xi), x(\tau,\xi)\big)
    \Big|_{\partial{\mycal C}_\epsilon^{-}}
    \\
    \label{eq:phi_L_cone}
    &=\varphi\Big(\tfrac1\epsilon - \tfrac12\big(\tfrac1\epsilon
    - \epsilon\big)
     e^{-\epsilon\tau - \epsilon\xi - \frac{\epsilon(1-\epsilon)}{1+\epsilon}},
     -\tfrac12\big(\tfrac1\epsilon - \epsilon\big)
     e^{-\epsilon\tau - \epsilon\xi
     - \frac{\epsilon(1-\epsilon)}{1+\epsilon}}\Big)
\end{align}
for $\xi \leq -\frac{1-\epsilon}{1+\epsilon}$. Equation
\eqref{eq:t=-x} implies that $X:=\partial_\tau=-\partial_\xi$, for
$\xi \geq \frac{1-\epsilon}{1+\epsilon}$, whereas \eqref{eq:t=x}
implies $X:=\partial_\tau=\partial_\xi$ for $\xi \leq
-\frac{1-\epsilon}{1+\epsilon}$. Within these regions of the
Cauchy surface, the dynamics consists in transporting the field
data $(\phi,\pi)$ over $\Sigma$ along the field $X$, according to
following equations:
\begin{align*}
    {\mycal L}_X \phi &= \partial_\tau  \phi = -\partial_\xi \phi
    \quad \text{for } \xi
    \geq
    \frac{1-\epsilon}{1+\epsilon},
    \\
    {\mycal L}_X\phi &= \partial_\tau  \phi = \partial_\xi \phi
    \quad \text{for } \xi \leq
    -\frac{1-\epsilon}{1+\epsilon},
    \\
    {\mycal L}_X\pi &= \partial_\tau  \pi = -\partial_\xi \pi
    \quad \text{for } \xi \geq
    \frac{1-\epsilon}{1+\epsilon},
    \\
    {\mycal L}_X\pi &= \partial_\tau  \pi = \partial_\xi \pi
    \quad \text{for } \xi \leq
    -\frac{1-\epsilon}{1+\epsilon},
\end{align*}
where ${\mycal L}_X$ denotes the Lie derivative along the vector
field $X$. The above equations can be derived from the following
Hamiltonians (generators of space translations):
\begin{align}
  \label{eq:H_[infty]}
  H_{[\frac{1-\epsilon}{1+\epsilon}, \infty)}(\phi , \pi ) &:=
  \int_{\frac{1-\epsilon}{1+\epsilon}}^{\infty}
  ( \pi \dot{\phi} -   \mathcal{L} ) \romand\xi
  = \int_{\frac{1-\epsilon}{1+\epsilon}}^{\infty}
  (-\pi \partial_{\xi} \phi) \romand\xi\; ,
\end{align}
\begin{align}
  \label{eq:H_[-infty]}
  H_{(-\infty,-\frac{1-\epsilon}{1+\epsilon}]}(\phi , \pi ) &:=
  \int^{-\frac{1-\epsilon}{1+\epsilon}}_{-\infty}
  ( \pi \dot{\phi} -   \mathcal{L} ) \romand\xi
  = \int^{-\frac{1-\epsilon}{1+\epsilon}}_{-\infty}
  (\pi \partial_{\xi} \phi) \romand\xi\; ,
\end{align}
where $\mathcal{L}$ vanishes identically as a pull-back of the scalar
density $L$ via the degenerate coordinate transformation
\eqref{eq:t=-x}-\eqref{eq:t=x}, and the momentum $\pi$ on the
Cauchy surface $\Sigma$ is equal to the pull-back of the (odd)
differential form $\pi^\mu \partial_{\mu} \rfloor \romand\xi^0\wedge
\romand\xi^1$ to $\partial {\mycal C}^{-}_{\epsilon}$. Moreover,
momentum $\pi^1$ coincides with $\pi=\pi^0$ for
$\xi\geq\frac{1-\epsilon}{1+\epsilon}$, and with $-\pi^0$ for
$\xi\leq -\frac{1-\epsilon}{1+\epsilon}$, as the pull-back of the
same form to the hypersurface
$\{\xi^1=\text{const.}\}=\{\xi^0=\text{const.}\}=\Sigma$. Hence,
we obtain the following constraints:
\begin{align}
  \label{eq:constrain>}
    \pi &= \dot{\phi} = - \partial_{\xi} \phi \quad \text{for }
    \xi\geq\frac{1-\epsilon}{1+\epsilon}\; ,
    \\
    \label{eq:constrain<}
    \pi &= \dot{\phi} =  \partial_{\xi} \phi \quad \text{for }
    \xi\leq-\frac{1-\epsilon}{1+\epsilon}\; .
\end{align}
The phase space of Cauchy data on the entire $\Sigma$ is described
by the pairs $(\phi, \pi)$ defined on the whole $\Real$ and
fulfilling constraint \eqref{eq:constrain>} or
\eqref{eq:constrain<} in the corresponding regions.

The total Hamiltonian function $H_\epsilon$ on the entire phase
space ${\mycal P}=\{(\phi,\pi)\}$ is equal to the sum of these
partial Hamiltonians:
\begin{align}
  \label{eq:H_e_total}
    H_\epsilon :=
    H_{(-\infty,-\frac{1-\epsilon}{1+\epsilon}]}+
    H_{[-\frac{1-\epsilon}{1+\epsilon}, \frac{1-\epsilon}{1+\epsilon}]}+
    H_{[\frac{1-\epsilon}{1+\epsilon}, \infty)} \ .
\end{align}
Variation of $H_\epsilon$ gives
\begin{align}
  \notag
    -\delta H_\epsilon(\phi,\pi) &= \int_\Sigma ({\mycal L}_X\pi \delta\phi
    - {\mycal L}_X\phi \delta\pi) \romand\xi
  \\
    \label{eq:dH}
    &+ \big[\pi^1\delta\phi\big]_{-\infty}^{-\frac{1-\epsilon}{1+\epsilon}}
    + \big[\pi^1\delta\phi\big]_{-\frac{1-\epsilon}{1+\epsilon}}
      ^{\frac{1-\epsilon}{1+\epsilon}}
    + \big[\pi^1\delta\phi\big]^{\infty}_{\frac{1-\epsilon}{1+\epsilon}}
\end{align}
with appropriate  values for $({\mycal L}_X\phi, {\mycal L}_X\pi)$
in the respective regions of $\Sigma$. The functional $H_\epsilon$
defines the Hamiltonian dynamics of the total system, if the
boundary terms in formula \eqref{eq:dH} cancel. This requires
corner conditions at $\xi=-\frac{1-\epsilon}{1+\epsilon}$ and
$\xi=\frac{1-\epsilon}{1+\epsilon}$ and sufficient strong fall-off
condition at infinity. To analyse these conditions it is useful to
reformulate our Hamiltonian description. Taking into account
constraints \eqref{eq:constrain>} and \eqref{eq:constrain<} and
using \eqref{eq:H_[infty]} and \eqref{eq:H_[-infty]} we have
\begin{align}
  \label{eq:H_[infty]_phi}
  H_{[\frac{1-\epsilon}{1+\epsilon}, \infty)} &=
  \int_{\frac{1-\epsilon}{1+\epsilon}}^{\infty} (\partial_{\xi} \phi)^2
  \romand\xi\ ,
  \\
  \label{eq:H_[-infty]_phi}
  H_{(-\infty,-\frac{1-\epsilon}{1+\epsilon}]} &=
  \int^{-\frac{1-\epsilon}{1+\epsilon}}_{-\infty}
  (\partial_{\xi} \phi)^2 \romand\xi\ ,
\end{align}
and the corresponding symplectic structures
\begin{align}
  \label{eq:omega_infty}
    \omega_{[\frac{1-\epsilon}{1+\epsilon}, \infty)}
    &=
    \int_{\frac{1-\epsilon}{1+\epsilon}}^{\infty}
    (-\partial_\xi \delta\phi \wedge \delta \phi) \romand\xi \quad \text{for }
    \xi\geq \frac{1-\epsilon}{1+\epsilon}\; ,
    \\
    \label{eq:omega_-infty}
    \omega_{(-\infty,-\frac{1-\epsilon}{1+\epsilon}]}
    &=
    \int^{-\frac{1-\epsilon}{1+\epsilon}}_{-\infty}
    (\partial_\xi \delta\phi \wedge \delta \phi) \romand\xi \quad \text{for }
    \xi\leq -\frac{1-\epsilon}{1+\epsilon}\; .
\end{align}
Changing variables in the following way:
\begin{align*}
    \lambda &:= \tau + \frac{1-\epsilon}{1+\epsilon}-\xi
      \quad \text{for } \xi \geq \frac{1-\epsilon}{1+\epsilon}\;,
    \\
    \chi &:= \tau + \frac{1-\epsilon}{1+\epsilon}+\xi
      \quad \text{for } \xi \leq -\frac{1-\epsilon}{1+\epsilon}\;,
\end{align*}
and denoting:
\begin{align*}
    x_{\epsilon}(\chi)
    &:= \phi\big(\tau, \chi-\tau-\tfrac{1-\epsilon}{1+\epsilon}\big)\;,
    \\
    y_{\epsilon}(\lambda)
    &:= \phi\big(\tau, \tau+\tfrac{1-\epsilon}{1+\epsilon}-\lambda\big)\;,
\end{align*}
we see that functions $x_{\epsilon}$ and $y_{\epsilon}$ do not
depend upon $\tau$ (see \eqref{eq:phi_R_cone} and
\eqref{eq:phi_L_cone}), hence they are single variable functions.
Now we can write formulas
\eqref{eq:H_[infty]_phi}-\eqref{eq:H_[-infty]_phi} and
\eqref{eq:omega_infty}-\eqref{eq:omega_-infty} jointly
\begin{align}
  \notag
    H_{\exter,\epsilon} &:= H_{(-\infty, -\frac{1-\epsilon}{1+\epsilon}]}
    + H_{[\frac{1-\epsilon}{1+\epsilon},\infty)}
    \\
    \label{eq:H_ext}
    &= \int_{-\infty}^{0} \big\{\big(\partial_{\lambda}
    f^{-}_{\epsilon}\big)^2
    + \big(\partial_{\lambda} g^{-}_{\epsilon}\big)^2 \big\}\romand\lambda
    \\
    \notag
    \omega_{\exter,\epsilon}
    &:= \omega_{(-\infty,-\frac{1-\epsilon}{1+\epsilon}]}
    +\omega_{[\frac{1-\epsilon}{1+\epsilon},\infty)}\; ,
    \\
    \label{eq:omega_ext}
    &= \int^{0}_{-\infty}
    \big\{ \partial_\lambda \delta f^{-}_{\epsilon}
      \wedge \delta f^{-}_{\epsilon}
    + \partial_\lambda \delta g^{-}_{\epsilon}
      \wedge \delta g^{-}_{\epsilon} \big\} \romand\lambda\;  ,
\end{align}
where
\begin{align}
  \label{eq:f(-)(e)}
    f^{-}_{\epsilon}(\tau,\lambda) &:= x_\epsilon(\lambda+\tau)\; ,
    \\
    \label{eq:g(-)(e)}
    g^{-}_{\epsilon}(\tau,\lambda) &:= y_\epsilon(\lambda+\tau)\; .
\end{align}
Formula \eqref{eq:omega_ext} shows, that the canonical structure
of ``external data'' (i.e.~data outside of the hyperboloid) can be
described by the ``$\int \delta f^\prime \wedge \delta f $''
symplectic form. To find the appropriate functional-analytic
framework for the problem and, in particular, to obtain correct
formulation of the corner condition, it is useful to reformulate
also the ``internal data'' (on the hyperboloid) in a similar way.
Indeed, we shall prove in the sequel that the transition between
the hyperboloidal data and that part of the light-cone data, which
lies {\em above the hyperboloid} is a canonical transformation,
which converts the canonical form ``$\int \delta \pi \wedge \delta
\phi $'' into the Faddeev form ``$\int \delta f^\prime \wedge
\delta f $'', (cf. \cite{Fad}).

For this purpose, assume that we know the light-cone data $(f, g)$, where
$f$ is a function which lives on the left piece of the light-cone, whereas
function $g$ lives on the right one.
We use convenient coordinates
\begin{align*}
  u &= t-x\;,\\
  v &= t+x\;,
\end{align*}
and in particular the left piece of $\Gamma_\epsilon :=\partial {\mycal
C}^{-}_{\epsilon}$ is given by
$\{u=\frac1\epsilon\}$, while the right one is given by
$\{v=\frac1\epsilon\}$. Therefore
\begin{align}
  \label{eq:f}
  \varphi\big(\tfrac12(v+\tfrac1\epsilon),\tfrac12(v-\tfrac1\epsilon)\big)
      &= \Phi\big(\tfrac1\epsilon\big) + \Psi(v)=f(v)\;,\\
    \label{eq:g}
    \varphi\big(\tfrac12(\tfrac1\epsilon+u),\tfrac12(\tfrac1\epsilon-u)\big)
      &= \Phi(u) + \Psi\big(\tfrac1\epsilon\big)=g(u)\;,
\end{align}
where $\varphi$ is the general solution of the wave equation,
i.e.
\begin{equation}\label{eq:gen_sol}
  \varphi (t,x)=\Phi(t-x) + \Psi(t+x)\;,
\end{equation}
where $\Phi$ and $\Psi$ are functions of one variable. Due to
\eqref{eq:f} and \eqref{eq:g}, we can express the general solution
\eqref{eq:gen_sol} in terms of the light-cone data $(f,g)$:
\begin{align}\label{eq:gen_sol_fg}
    \varphi (t,x)=f(t+x)+ g(t-x) - \varphi\big(\tfrac1\epsilon,0\big)\;,
\end{align}
where
$\varphi\big(\frac1\epsilon,0\big)=\Phi\big(\frac1\epsilon\big)
+\Psi\big(\frac1\epsilon\big)=f\big(\frac1\epsilon\big)
=g\big(\frac1\epsilon\big)$. Formula \eqref{eq:gen_sol_fg} implies
the following transformation between the hyperboloidal data
$(\phi, \pi)$ and the light-cone data $(f,g)$:
\begin{align}
  \notag
    \phi(\tau,\xi) &= \varphi\big|_{\tau=\text{const.}}
    \\
  \label{eq:tr_phi}
    &=f\Big(\tfrac{1}{\epsilon}+
      \big(\tfrac{1+\xi}{1-\xi}-\tfrac{1}{\epsilon}\big)e^{-\epsilon\tau}\Big)
      +g\Big(\tfrac{1}{\epsilon}+
      \big(\tfrac{1-\xi}{1+\xi}-\tfrac{1}{\epsilon}\big)e^{-\epsilon\tau}\Big)
      -\varphi\big(\tfrac1\epsilon,0\big)\;,
    \\
    \notag
    \pi(\tau,\xi) &= \kappa^{-1} (\partial_\tau \varphi
      + \xi\partial_\xi \varphi)\big|_{\tau=\text{const.}}
    \\
    \label{eq:tr_pi}
    &=\frac{2e^{-\epsilon\tau}}{(1-\xi)^2}
    f'\Big(\tfrac{1}{\epsilon}+
      \big(\tfrac{1+\xi}{1-\xi}-\tfrac{1}{\epsilon}\big)e^{-\epsilon\tau}\Big)
    +\frac{2e^{-\epsilon\tau}}{(1+\xi)^2}
    g'\Big(\tfrac{1}{\epsilon}+
      \big(\tfrac{1-\xi}{1+\xi}-\tfrac{1}{\epsilon}\big)e^{-\epsilon\tau}\Big)\;,
\end{align}
where we used definition \eqref{eq:pi^0} of the momentum $\pi$.
Integrating \eqref{eq:tr_pi} over intervals
$[-\frac{1-\epsilon}{1+\epsilon},\xi]$ and
$[\xi,\frac{1-\epsilon}{1+\epsilon}]$, we obtain, due to
\eqref{eq:tr_phi}, the following transformation:
\begin{align}
  \label{eq:tr_f}
    f(v) &= \frac 1 2 \Big\{
    \phi\Big(-\tfrac{1-ve^{\epsilon\tau}-\frac1\epsilon(1-e^{\epsilon\tau})}
    {1+ve^{\epsilon\tau}+\frac1\epsilon(1-e^{\epsilon\tau})}\Big)
    +\phi\big(-\tfrac{1-\epsilon}{1+\epsilon}\big)\Big\}
    + \frac 1 2 \int_{-\frac{1-\epsilon}{1+\epsilon}}^
      {-\frac{1-ve^{\epsilon\tau}-\frac1\epsilon(1-e^{\epsilon\tau})}
    {1+ve^{\epsilon\tau}+\frac1\epsilon(1-e^{\epsilon\tau})}}
    \pi(\eta) \romand \eta\;,
\end{align}
and
\begin{align}
    \label{eq:tr_g}
    g(u) &= \frac 1 2 \Big\{ \phi\big(\tfrac{1-\epsilon}{1+\epsilon}\big)
    +\phi\Big(\tfrac{1-ue^{\epsilon\tau}-\frac1\epsilon(1-e^{\epsilon\tau})}
    {1+ue^{\epsilon\tau}+\frac1\epsilon(1-e^{\epsilon\tau})}\Big) \Big\}
    + \frac 1 2
    \int^{\frac{1-\epsilon}{1+\epsilon}}_
    {\frac{1-ue^{\epsilon\tau}-\frac1\epsilon(1-e^{\epsilon\tau})}
    {1+ue^{\epsilon\tau}+\frac1\epsilon(1-e^{\epsilon\tau})}}
    \pi(\eta) \romand \eta\;.
\end{align}

Substitution of \eqref{eq:tr_phi}, \eqref{eq:tr_pi} into the
symplectic form \eqref{eq:symp_phi_pi} defined on the space of
Cauchy data $(\phi,\pi)$ over the hyperboloid $V_\epsilon$ gives
us:
\begin{multline}
  \label{eq:df'df+dg'dg}
  \omega_{\inter,\epsilon}=
    \int_{-\frac{1-\epsilon}{1+\epsilon}}^{\frac{1-\epsilon}{1+\epsilon}}
    \delta \pi(\tau,\xi) \wedge \delta \phi(\tau,\xi) \romand \xi
    \\
    = \int_{\frac{1}{\epsilon}
    +(\epsilon-\frac{1}{\epsilon})e^{-\epsilon\tau}}^{\frac{1}{\epsilon}}
    \left\{
    \partial_z\delta f(z) \wedge \delta f(z)
    +
    \partial_z\delta g(z)   \wedge \delta g(z)
    \right\}\romand z\;.
\end{multline}
Changing variables in \eqref{eq:df'df+dg'dg} in the following way
\begin{align*}
    z = {\frac{1}{\epsilon}} +\Big(\epsilon-\frac{1}{\epsilon}\Big)
    e^{-\epsilon\lambda-\epsilon\tau}
\end{align*}
and denoting functions
\begin{align}
  \label{eq:f(+)(e)}
    f^{+}_{\epsilon}(\tau,\lambda) &:= f\Big({\tfrac{1}{\epsilon}}
    +\big(\epsilon-\tfrac{1}{\epsilon}\big)
    e^{-\epsilon\lambda-\epsilon\tau}\Big)\;,
    \\
    \label{eq:g(+)(e)}
    g^{+}_{\epsilon}(\tau,\lambda) &:= g\Big({\tfrac{1}{\epsilon}}
       +\big(\epsilon-\tfrac{1}{\epsilon}\big)
       e^{-\epsilon\lambda-\epsilon\tau}\Big)\;,
\end{align}
(we use superscript "$+$", because these functions are defined
over the positive half-line $\lambda\in[0,\infty)$) we can write
the formula \eqref{eq:df'df+dg'dg} in the following form
\begin{align}
   \label{eq:omega_ext_e=omega[0,infty]}
   \omega_{\inter,\epsilon}
    &= \int_{0}^{\infty}
    \big\{ \partial_\lambda \delta f^{+}_{\epsilon}
      \wedge \delta f^{+}_{\epsilon}
    + \partial_\lambda \delta g^{+}_{\epsilon}
      \wedge \delta g^{+}_{\epsilon}\big\} \romand\lambda\ .
\end{align}
Equation \eqref{eq:omega_ext_e=omega[0,infty]} shows that, indeed,
the "internal" part of the canonical structure can also be
described by the ``$\int \delta f^\prime \wedge \delta f $''
symplectic form defined on the light-cone data. This ends our
proof.

Now, the corner conditions at points $\xi=\pm
\frac{1-\epsilon}{1+\epsilon}$ (necessary for cancellation of
boundary terms in the Hamiltonian formula \eqref{eq:dH}) are
expressed into the compatibility condition between the external
(given by \eqref{eq:omega_ext}) and the internal (given by
\eqref{eq:omega_ext_e=omega[0,infty]}) structures, which must be
satisfied at the point $\lambda = 0$. An obvious condition is that
the total symplectic form:
\begin{align}
    \omega_\epsilon:= \omega_{\exter,\epsilon}+\omega_{\inter,\epsilon}
    =\int_{-\infty}^{\infty} \left\{
     \delta f_\epsilon^\prime \wedge \delta f_\epsilon
    +
    \delta g_\epsilon^\prime  \wedge \delta g_\epsilon
    \right\}\romand \lambda \ ,
\end{align}
where $f_\epsilon$, $g_\epsilon$ are equal to $f^{-}_\epsilon$,
$g^{-}_\epsilon$ for $\lambda < 0$  and to $f^{+}_\epsilon$,
$g^{+}_{\epsilon}$ for $\lambda > 0$, respectively, must be well
defined. In particular, a step discontinuity is excluded, because
its derivative would produce the Dirac delta, which cannot be
integrated with a non-continuous function. We see that, due to
constrains \eqref{eq:constrain>}--\eqref{eq:constrain<}, the
Faddeev-Takhtajan-Dubrovin symplectic form, typical for the
Hamiltonian theory of solitons, arises in a natural way in
Hamiltonian description of characteristic value problem for
standard, hyperbolic equations. Transition between the standard
Cauchy data and the radiation (characteristic) data is a
symplectomorphism.

We conclude that the product of $f_\epsilon$ and $g_\epsilon$ by
their derivatives must be integrable on real line $\mathbb{R}$, so
must belong to $\Lspace^1(\Real)$. Moreover, functions
$f_\epsilon$, $g_\epsilon$ and $f_\epsilon'$, $g_\epsilon'$ have
to belong to mutually dual spaces because they
represent``positions" and ``momenta" correspondingly. This
implies: that $f_\epsilon, g_\epsilon \in \Hspace^{\frac 1
2}(\Real)$ and, then, $f_\epsilon', g_\epsilon' \in
\Hspace^{-\frac 1 2}(\Real)$. Equations
\eqref{eq:tr_phi}--\eqref{eq:tr_pi} imply that we have also: $\phi
\in \Hspace^{\frac 1 2}(\Real)$ and $\pi \in \Hspace^{-\frac 1
2}(\Real)$.

Due to equations  \eqref{eq:f(-)(e)}--\eqref{eq:g(-)(e)} and
\eqref{eq:f(+)(e)}--\eqref{eq:g(+)(e)}), we have the following
relation between the previous and the final representation of
Cauchy data:
\begin{align*}
    f^{-}_\epsilon (\tau,\lambda) &=
    \phi(\tau, \lambda-\tfrac{1-\epsilon}{1+\epsilon})\ ,
    \\
    g^{-}_\epsilon (\tau,\lambda) &=
    \phi(\tau,\tfrac{1-\epsilon}{1+\epsilon}-\lambda)\ ,
\end{align*}
and
\begin{align*}
    f^{+}_\epsilon (\tau,\lambda) &=
    \frac 1 2 \Big\{
    \phi\Big(\tau, -\tfrac{(1-\epsilon)(1+\epsilon-e^{\epsilon\lambda})}
    {(1+\epsilon)(-1+\epsilon+e^{\epsilon\lambda})}\Big)
    +\phi\big(\tau, -\tfrac{1-\epsilon}{1+\epsilon}\big)\Big\}
    + \frac 1 2 \int_{-\frac{1-\epsilon}{1+\epsilon}}^
      {-\tfrac{(1-\epsilon)(1+\epsilon-e^{\epsilon\lambda})}
    {(1+\epsilon)(-1+\epsilon+e^{\epsilon\lambda})}}
    \pi(\tau,\xi) \romand \xi\ ,
    \\
    g^{+}_\epsilon (\tau,\lambda) &=
    \frac 1 2 \Big\{ \phi\big(\tau, \tfrac{1-\epsilon}{1+\epsilon}\big)
    +\phi\Big(\tau, \tfrac{(1-\epsilon)(1+\epsilon-e^{\epsilon\lambda})}
    {(1+\epsilon)(-1+\epsilon+e^{\epsilon\lambda})}\Big) \Big\}
    + \frac 1 2 \int^{\frac{1-\epsilon}{1+\epsilon}}_
      {\tfrac{(1-\epsilon)(1+\epsilon-e^{\epsilon\lambda})}
    {(1+\epsilon)(-1+\epsilon+e^{\epsilon\lambda})}}
    \pi(\tau, \xi) \romand \xi\ .
\end{align*}

The Hamiltonian system we have obtained is conservative, because
the total Hamiltonian \eqref{eq:H_e_total} does not depend
explicitly on time. Hence, the total energy is conserved:
\begin{align*}
    \frac{\romand}{\romand\tau}H_\epsilon=0\ .
\end{align*}
But, due to \eqref{eq:H_[infty]_phi} and
\eqref{eq:H_[-infty]_phi}, the external energy
$H_{(-\infty,-\frac{1-\epsilon}{1+\epsilon}]}$ and
$H_{[\frac{1-\epsilon}{1+\epsilon}, \infty)}$ are monotonically
increasing function on time. Indeed, they are obtained by
integration of a non-negative function $(\partial_\xi \phi)^2$
over a part of the boundary $\partial {\mycal C}^{-}_{\epsilon}$
of the cone which increases when the time increases. We conclude
that the Trautman-Bondi internal
$H_{[-\frac{1-\epsilon}{1+\epsilon},
\frac{1-\epsilon}{1+\epsilon}]}$ on the hyperboloid must be
monotonically decreasing function on time.

We have also showed that $H_{[-\frac{1-\epsilon}{1+\epsilon},
\frac{1-\epsilon}{1+\epsilon}]}$ is positive inside the cone. It
represents the amount of energy still remaining in the system,
whereas $H_{(-\infty,-\frac{1-\epsilon}{1+\epsilon}]}$ and
$H_{[\frac{1-\epsilon}{1+\epsilon}, \infty)}$ describe already
radiated energy.

\subsection{Four-dimensional Minkowski spacetime}
\label{sec:13case}

The above construction, with appropriate modifications, is valid
also in four-dimensional Minkowski spacetime. Consider a
one-parameter family of past-oriented light cones:
\begin{equation*}
  {\mycal C}_\epsilon^-:=
  \left\{ (t,x):\; x \in {\mathbb R}^3\, ,\;
  \frac1\epsilon - t > \|x\| \right\}\;,
\end{equation*}
where $\frac1\epsilon >1$ is the time coordinate of the vertex of
${\mycal C}_\epsilon^-$. We introduce  new coordinates $(\xi^\mu)
= (\tau , \xi^k )$ ($\mu = 0,\ldots, 3$), related to Minkowskian
coordinates $(x^\mu) = (t , x^k)$ in a way analogous to
\eqref{eq:t_xi}-\eqref{eq:x_xi}:
\begin{align}
  \label{eq:t_xi_4d}
  t &=  \frac1\epsilon + \left(\frac{1+\|\xi\|^2}{1-\|\xi\|^2}
  -\frac1\epsilon \right)e^{-\epsilon\tau} \;,
  \\
  \label{eq:x_xi_4d}
  x^k &=  \frac{2\xi^k}{1-\|\xi\|^2} e^{-\epsilon\tau} \;,
\end{align}
where $\tau \in {\mathbb R}^1$. For $\| \xi \| \leq
\frac{1-\epsilon}{1+\epsilon}$, the new coordinates parameterize
the entire cone ${\mycal C}_\epsilon^-$ and the surfaces $\{\tau=
\text{const.}\}$ correspond to hyperboloids. To derive the
Hamiltonian description of the wave equation in these coordinates,
we begin with the Lagrangian:
\begin{equation}\label{Lagr-st_4d_we}
  \hat{L} = - \frac 12 \sqrt{|\det g |}
  g^{\mu\nu} (\partial_\mu \hat{\varphi} )(\partial_\nu \hat{\varphi} ) =
  \frac 12 \left\{ \left( \partial_t \hat{\varphi} \right)^2 -
  \left( \nabla \hat{\varphi} \right)^2 \right\} \;.
\end{equation}
Expressing it in terms of new coordinates we obtain:
\begin{equation}
    \label{eq:Lagr-st_4d}
  \hat{L} \cdot \textrm{d}^4x = \hat{\mathcal{L}} \cdot \textrm{d}^4\xi \;,
\end{equation}
where
\begin{equation}
  \hat{\mathcal{L}}  =
  \left\{ \frac{\left( \frac{\partial \hat\varphi}{\partial\tau} + \xi^k
  \frac{\partial \hat\varphi}{\partial\xi^k}\right)^2}
  {1-\epsilon+(1+\epsilon)\|\xi\|^2}
  - \frac14\left[{1-\epsilon+(1+\epsilon)\|\xi\|^2} \right] \delta^{kl}
   \frac{\partial \hat\varphi}{\partial\xi^k}
   \frac{\partial \hat\varphi}{\partial\xi^l}
   \right\} \left( \frac{2e^{-\epsilon\tau}}{1-\|\xi\|^2}\right)^2
\, .
\end{equation}
This, apparently non-autonomous (i.e. $\tau$-dependent),
Lagrangian becomes autonomous after an appropriate re-scaling of
the field variable:
\begin{equation}
  \label{eq:hat_varphi->varphi}
  \varphi:=  \frac{2e^{-\epsilon\tau}}{1-\|\xi\|^2} \hat\varphi \ .
\end{equation}
Indeed, we obtain the formula analogous to \eqref{eq:Lagr-st-xi}:
\begin{align}
  \notag
  \hat{\mathcal{L}} &= \frac1{2\kappa} \left[ \partial_\tau \varphi
  +\xi^k\partial_{\xi^k}\varphi
  +\left(\epsilon -\frac{2\|\xi\|^2}{1-\|\xi\|^2}\right)\varphi \right]^2
  \\
  &- \frac12 \kappa  \delta^{kl}\left( \partial_{\xi^k}\varphi
  -\frac{2\xi_k}{1-\|\xi\|^2}\varphi\right) \left( \partial_{\xi^l}\varphi
  -\frac{2\xi_l}{1-\|\xi\|^2}\varphi\right) \, ,
\end{align}
where we denote:
\[
  \kappa:= \frac12 \left[ 1-\epsilon+(1+\epsilon)\|\xi\|^2 \right] \, .
\]
Observe that the function
\begin{multline}
  \label{eq:Lagr-ren}
  \mathcal{L} = \frac1{2\kappa} \left( \partial_\tau\varphi
  +\xi^k \partial_{\xi^k} \varphi
   \right)^2 - \frac12 \kappa
   \delta^{kl}(\partial_{\xi^k}\varphi)(\partial_{\xi^l}\varphi)
  \\
  - \frac{(1-\epsilon)(3+\epsilon^2 )
  - 2(1-\epsilon)(1+\epsilon)^2 \|\xi\|^2 - (1+\epsilon)^3\|\xi\|^4}
  {2\left(1-\epsilon+\left(1+\epsilon\right)\|\xi\|^2  \right)^2} \varphi^2\;,
\end{multline}   
differs from the original Lagrangian $\hat{\mathcal{L}}$  by a complete
divergence:
\begin{equation*}
  \hat{\mathcal{L}} = \mathcal{L} + \partial_\mu \mathcal{Z}^\mu
  \, ,
\end{equation*}
where
\begin{equation*}
  \mathcal{Z}^0 = \frac1{2\kappa}
  \left(\epsilon -\frac{2\|\xi\|^2}{1-\|\xi\|^2}\right)\varphi^2 \, ,
\end{equation*}
and
\begin{equation*}
  \mathcal{Z}^k = \frac1{4\kappa}\xi^k
  \left[ 1+\epsilon^2 -(1+\epsilon)^2\|\xi\|^2 \right]\varphi^2 \, .
\end{equation*}
This implies that both Lagrangians $\hat{\mathcal{L}}$ and $\mathcal{L}$
lead to the same equation of motion for the field $\varphi$, so we
use the latter in the sequel.

To derive the Hamiltonian description, we integrate equation
\eqref{eq:delta-cL1} over the volume $V_{\inter, \epsilon}:=\{\xi
: \|\xi\|\leq \frac{1-\epsilon}{1+\epsilon}\}$ in the Cauchy
surface $\Sigma =\{\tau=\text{const.}\}$. We obtain an identity
valid for fields satisfying wave equation:
\begin{align*}
    \delta \int_{V_{\inter,\epsilon}} \mathcal{L} \romand^3\xi
    &= \int_{V_{\inter,\epsilon}}
    \left( \pi \delta \phi \right)^{\cdot}\romand^3\xi
    + \int_{\partial V_{\inter,\epsilon}}
    \pi^k \delta \phi\;  \romand^2\sigma_k\;,
\end{align*}
where "dot" denotes derivative with respect to the new time
variable $\tau$, while $\phi$ is the restriction of the field
$\varphi$ to the Cauchy surface $\Sigma=\{\tau=\text{const.}\}$.
Moreover, we have introduced momentum $\pi:=\pi^0$, which is a
part of Cauchy data on this surface. Performing Legendre
transformation between $\dot{\phi}$ and $\pi$ we get:
\begin{align}
  \label{eq:dH_int_e}
    -\delta H_{\inter,\epsilon}
    = \int_{V_{\inter,\epsilon}}
    ( \dot{\pi}\delta \phi-\dot{\phi}\delta \pi) \romand^3\xi
    + \int_{\partial V_{\inter,\epsilon}} \pi^{\bot} \delta \phi\;,
\end{align}
where the Hamiltonian $H_{{\inter,\epsilon}}$ is defined by
\begin{align}
  \label{eq:H_int_e}
  H_{{\inter,\epsilon}}(\phi , \pi ) &:=  \int_{V_{\inter,\epsilon}}
  \big( \pi \dot{\phi} -   \mathcal{L} \big) \romand^3\xi\;,
\end{align}
and the symplectic form in phase space of Cauchy data is given by:
\begin{align}\label{eq:symp_int_e}
    \omega_{{\inter,\epsilon}} &:= \int_{V_{\inter,\epsilon}}
    (\delta\pi\wedge\delta\phi)\romand^3\xi\;.
\end{align}
The Lagrangian \eqref{eq:Lagr-ren} implies the following relation
between ``velocity'' $\dot{\phi}$ and momentum~$\pi$:
\begin{align}
    \pi &= \pi^0 =
    \frac {\partial \mathcal{L}}{\partial\varphi_0} =
  \kappa^{-1} \left(\partial_{\tau}\varphi + \xi^k
  \frac {\partial \varphi}{\partial \xi^k} \right)
  = \kappa^{-1} \left(\dot{\phi} + \xi^k
  \frac {\partial \phi}{\partial \xi^k} \right)\, .
\end{align}
Thus, the Hamiltonian \eqref{eq:H_int_e} my be written explicitly
in terms of the Cauchy data on $V_{\inter,\epsilon}$
\begin{align}
  \label{eq:H(phi,pi)_int_e}
   H_{{\inter,\epsilon}}(\phi , \pi )
  &=
  \frac 12 \int_{V_{\inter,\epsilon}}  \left\{ \kappa \left( \pi
  -\kappa^{-1}\xi^l
  \frac {\partial \phi}{\partial \xi^l}\right)^2 +
  \left(\kappa\delta^{kl}-\kappa^{-1}\xi^k
   \xi^l \right)\frac{\partial \phi}{\partial \xi^k}
  \frac{\partial \phi}{\partial \xi^l}  +\mu \phi^2
    \right\} \romand^3\xi\;,
\end{align}
where
\begin{equation*}
  \mu:= \frac{(1-\epsilon)(3+\epsilon^2 )
  - 2(1-\epsilon)(1+\epsilon)^2 \|\xi\|^2 - (1+\epsilon)^3\|\xi\|^4}
  {\left(1-\epsilon+\left(1+\epsilon\right)\|\xi\|^2  \right)^2} \, .
\end{equation*}
One can check that the quadratic form
\begin{align*}
    \kappa\delta^{kl}-\kappa^{-1}\xi^k\xi^l
\end{align*}
is positive definite for $\kappa-\|\xi\|>0$. But, we have:
\begin{align*}
    \kappa-\|\xi\|=\frac12
    (1+\epsilon)(1-\|\xi\|)\bigg(\frac{1-\epsilon}{1+\epsilon}-\|\xi\|\bigg)
    \ .
\end{align*}
This implies that Hamiltonian \eqref{eq:H(phi,pi)_int_e} is
positive for $\|\xi\|<\frac{1-\epsilon}{1+\epsilon}$, i.e.~inside
the cone $\mycal{ C}_\epsilon^-$. The Euler-Lagrange equation
coincides with the following Hamiltonian equations, derived
directly from \eqref{eq:H(phi,pi)_int_e}:
\begin{align}
  \label{eq:dot_phi_int}
  \dot{\phi} &= \kappa\pi -\xi^k
  \frac {\partial \phi}{\partial \xi^k} \, ,\\
  \label{eq:dot_pi_int}
  \dot\pi &= -\partial_{\xi^k} (\xi^k\pi) + \partial_{\xi^k}(\kappa
  \delta^{kl}\partial_{\xi^l}\phi)-\mu\phi\, ,
\end{align}
provided no boundary terms remain after the integration by part of
its variation. To assure their cancellation we proceed in a way
analogous to the previous section: we take into account missing
radiation data on the light cone. Therefore, we extend
parametrization \eqref{eq:t_xi_4d}--\eqref{eq:x_xi_4d} beyond the
volume $V_{\inter,\epsilon}$, taking into account also the
corresponding points on the boundary of the cone:
\begin{align}
  \label{eq:t_V_ext}
    t &:= \frac1\epsilon - \frac12\Big(\frac1\epsilon - \epsilon\Big)
    e^{-\epsilon\tau + \epsilon\|\xi\|
    - \frac{\epsilon(1-\epsilon)}{1+\epsilon}} \quad
     \text{for}\ \|\xi\|\geq\frac{1-\epsilon}{1+\epsilon} \;,
    \\
    \label{eq:x_V_ext}
    x^k &:= \frac12\Big(\frac1\epsilon - \epsilon\Big)
     \frac{\xi^k}{\|\xi\|}e^{-\epsilon\tau + \epsilon\|\xi\|
     - \frac{\epsilon(1-\epsilon)}{1+\epsilon}} \quad
     \text{for}\ \|\xi\| \geq \frac{1-\epsilon}{1+\epsilon} \;,
\end{align}
and consider the data $(\phi,\pi)$ on the entire surface $\Sigma =
\{\tau=\text{const.},\xi\in \Real^3\}$. Equations
\eqref{eq:t_V_ext}--\eqref{eq:x_V_ext} imply that
$X:=\partial_\tau=-\frac{\xi^k}{\|\xi\|}\partial_{\xi^k}$, for
$\|\xi\| \geq \frac{1-\epsilon}{1+\epsilon}$. The dynamics
consists in transporting the field data $(\phi,\pi)$ over the
surface $\Sigma$ according to the following field equations:
\begin{align}
  \label{eq:dot_phi_ext}
    {\mycal L}_X \phi &= \partial_\tau  \phi
    = -\frac{\xi^k}{\|\xi\|}\partial_{\xi^k} \phi\;,
    \\
    \label{eq:dot_pi_ext}
    {\mycal L}_X\pi &= \partial_\tau  \pi = -\partial_{\xi^k}
    \left(\frac{\xi^k}{\|\xi\|}\pi\right)\;,
\end{align}
where \eqref{eq:dot_pi_ext} follows from the fact that the
momentum $\pi$ is not a scalar field (like $\phi$) but a scalar
density. The above equations can be also derived from the standard
Hamiltonian formula:
\begin{align}
  \label{eq:H_ext_e}
  H_{{\exter,\epsilon}}(\phi , \pi ) &:=
  \int_{V_{\exter,\epsilon}}
  ( \pi \dot{\phi} -   \mathcal{L} ) \romand^3\xi
  = \int_{V_{\exter,\epsilon}}
  \left(-\pi \frac{\xi^k}{\|\xi\|}\partial_{\xi^k} \phi\right) \romand^3\xi\;,
\end{align}
where $V_{\exter,\epsilon}:=\big\{\xi: \|\xi\| \geq
\frac{1-\epsilon}{1+\epsilon}\big\}$ and $\mathcal{L}$ vanishes
identically as a pull-back of the scalar density $L$ via the
degenerate coordinate transformation
\eqref{eq:t_V_ext}--\eqref{eq:x_V_ext}. Variation of the above
Hamiltonian gives:
\begin{align}
  -\delta H_{{\exter,\epsilon}}(\phi , \pi ) &=
  \int_{V_{\exter,\epsilon}}
  ( \dot{\pi} \delta{\phi} -  \dot{\phi} \delta{\pi} )\; \romand^3\xi
  + \int_{\partial V_{\exter,\epsilon}} \pi^{\bot}\delta\phi\; ,
\end{align}
where $(\dot{\phi},\dot{\pi})$ are given by \eqref{eq:dot_phi_ext}
and \eqref{eq:dot_pi_ext}, whereas the boundary term comes from
integration by parts. The momentum $\pi$ on the Cauchy surface
$\Sigma$ is equal to the pull-back of the differential (odd) form
$\pi^\mu \partial_{\mu} \rfloor \romand\xi^0\wedge
\romand\xi^1\wedge \romand\xi^2\wedge \romand\xi^3$ to
$\partial {\mycal C}^{-}_{\epsilon}$. Moreover,
momentum $\pi^{\bot}$ coincides with $\pi$, as the pull-back of
the same form to the the hypersurface
$\{\|\xi\|=\text{const.}\}=\{\tau=\text{const.}\}=\Sigma$, so we
obtain the following constraints
\begin{align}
  \label{eq:constrain_ext}
    \pi &=  - \frac{\big(\frac{1-\epsilon}{1+\epsilon}\big)^2}{\|\xi\|^2}
    \frac{\xi^k}{\|\xi\|}\partial_{\xi^k} \phi
    \quad\text{ for } \|\xi\| \geq \frac{1-\epsilon}{1+\epsilon} \ .
\end{align}
The phase space of Cauchy data on the entire $\Sigma$ is described
by the pairs $(\phi, \pi)$ defined on the whole $\Real^3$
fulfilling constraints \eqref{eq:constrain_ext} outside of the
hyperboloid. Moreover, functions $\phi$ and $\pi$ should satisfy
compatibility conditions ("corner conditions") at points $\|\xi\|
= \frac{1-\epsilon}{1+\epsilon}$, because otherwise the total
dynamics is not well defined. To formulate these conditions we
proceed as in the previous section. Summing up formula
\eqref{eq:H_int_e} for $H_{\inter,\epsilon}$ and formula
\eqref{eq:H_ext_e} for $H_{\inter,\epsilon}$ we define the total
energy $H_\epsilon$, defined on the total phase space ${\mycal
P}=\{(\phi,\pi)\}$
\begin{align*}
    H_\epsilon := H_{\inter,\epsilon}+H_{\exter,\epsilon}\; .
\end{align*}
Variation of $H_\epsilon$ gives us:
\begin{align}
  \label{eq:dH_4d}
    -\delta H_\epsilon(\phi,\pi) &= \int_\Sigma ({\mycal L}_X\pi \delta\phi
    - {\mycal L}_X\phi \delta\pi)\; \romand^3\xi
    + \int_{\partial V_{\inter,\epsilon}} \pi^{\bot} \delta \phi
    + \int_{\partial V_{\exter,\epsilon}} \pi^{\bot} \delta \phi\; ,
\end{align}
where the dynamics $({\mycal L}_X\phi, {\mycal L}_X\pi)$ is given
given by \eqref{eq:dot_phi_int}--\eqref{eq:dot_pi_int} (inside)
and by \eqref{eq:dot_phi_ext}--\eqref{eq:dot_pi_ext} outside of
the sphere $\|\xi\| = \frac{1-\epsilon}{1+\epsilon}$. The global
dynamics generated by $H_\epsilon$ is well defined if the boundary
terms in formula \eqref{eq:dH_4d} vanish. To analyze the resulting
corner conditions we reformulate our Hamiltonian description as
follows. Taking into account constraint \eqref{eq:constrain_ext}
we obtain from \eqref{eq:H_ext_e}:
\begin{align}
  \notag
  H_{\exter,\epsilon} &=
  \int_{V_{\exter,\epsilon}}
  \frac{\big(\frac{1-\epsilon}{1+\epsilon}\big)^2}{\|\xi\|^2}
  \left( \frac{\xi^k}{\|\xi\|}\partial_{\xi^k} \phi \right)^2 \romand^3\xi
  \\
  &=\int_{S^2}\int_{\rho \geq \frac{1-\epsilon}{1+\epsilon}}
  \left( \partial_{\rho} \phi\right)^2
  \big(\tfrac{1-\epsilon}{1+\epsilon}\big)^2
  \romand\rho \romand^2\sigma\;,
\end{align}
and the corresponding symplectic structure
\begin{align}
  \notag
    \omega_{\exter,\epsilon} &= -\int_{V_\exter,\epsilon}
    \frac{\big(\frac{1-\epsilon}{1+\epsilon}\big)^2}{\|\xi\|^2}
    \left( \frac{\xi^k}{\|\xi\|}\partial_{\xi^k} \delta\phi\right)
    \wedge \delta\phi\; \romand^3\xi
    \\
    \label{eq:int_rho>1}
    &=-\int_{S^2}\int_{\rho \geq \frac{1-\epsilon}{1+\epsilon}}
  \left( \partial_{\rho} \delta\phi\wedge \delta\phi\right)
  \big(\tfrac{1-\epsilon}{1+\epsilon}\big)^2  \romand\rho \romand^2\sigma\;,
\end{align}
where $\rho=\|\xi\|$ and $\romand^2\sigma$ denotes the volume
element on the two-dimensional unit sphere $S^2:=\{\xi\in\Real^3:
\|\xi\|=1\}$. Changing variables in the integral
\eqref{eq:int_rho>1}
\begin{align*}
    \rho=\tau + \frac{1-\epsilon}{1+\epsilon}-\lambda\;,
\end{align*}
the remaining variables being unchanged, and denoting
\begin{align*}
  y_\epsilon(\lambda,\dots)
  :=\phi(\tau + \tfrac{1-\epsilon}{1+\epsilon}-\lambda, \dots)
\end{align*}
we obtain that $y_\epsilon$ does not depend on variable $\tau$
(see formulas \eqref{eq:t_V_ext} and \eqref{eq:x_V_ext}). Hence,
we have:
\begin{align}
  \label{eq:int_lambda<0}
  \omega_{\exter,\epsilon}
  &= \int_{S^2}\int_{\lambda \leq 0}
  \left( \partial_{\lambda} \delta f^{-}_\epsilon
  \wedge \delta f^{-}_\epsilon \right)
  \romand\lambda \romand^2\sigma\; ,
\end{align}
where
\begin{align*}
    f^{-}_{\epsilon}(\lambda,\dots)
    :=\frac{1-\epsilon}{1+\epsilon}\ y_\epsilon(\lambda +\tau,\dots)\; .
\end{align*}
Expression \eqref{eq:int_lambda<0} for ``external" symplectic form
suggests to consider, instead of the phase space ${\mycal P}=
\{(\phi,\pi)\}$ with constraint \eqref{eq:constrain_ext}, the
phase space of functions defined on the half of the tube $\Real
\times S^2$, corresponding to negative values of $\lambda \in
\Real$. We will show in the sequel that ``internal'' data
$(\phi,\pi)$ on the hyperboloid can also be represented by
canonically equivalent data on the remaining half-tube. The proof
is based on the Euler-Lagrange equations, equivalent to the
following identity:
\begin{align}
  \label{eq:Eu-Lag_p}
    \delta \mathcal{L} = \partial_{\mu}(p^\mu \delta \varphi)\; ,
\end{align}
where by $p^\mu$ we denote generalized momenta. Integrating
equation \eqref{eq:Eu-Lag_p} over any region $\mathcal{V}$ in
spacetime we obtain identity:
\begin{align}
    \delta \int_{\mathcal{V}} \mathcal{L} =
    \int_{\partial\mathcal{V}}p^{\bot}\delta\varphi\; ,
\end{align}
which holds for any configuration $\varphi$ satisfying the field
equations. In particular, let $\mathcal{V}$ be the set of points
lying between the boundary of the cone
$\Gamma_\epsilon:=\partial{\mycal C}_\epsilon^{-}$ and the
hyperboloid $V_{\inter,\epsilon}$, then
\begin{align}
    \delta \int_{\mathcal{V}} \mathcal{L} =
    \int_{\Gamma_\epsilon} p^{\bot}\delta\varphi - \int_{V_{\inter,\epsilon}}
    p^{\bot}\delta\varphi\; ,
\end{align}
where the signs come from the orientation of both surfaces
$\Gamma_\epsilon$ and $V_{\inter, \epsilon}$. Now, we treat these
expressions as exterior one-forms on the space of Cauchy data on
$\Gamma_\epsilon$ and $V_{\inter, \epsilon}$ respectively, and
calculate exterior derivative of both sides. Because the
left-hand-side is already an exterior derivative, its further
exterior differentiation gives zero. This way we prove the
identity:
\begin{align}
  \label{eq:symp_Gamma=symp_V_int}
    \int_{V_{\inter,\epsilon}} \delta p^{\bot}\wedge\delta\varphi
    = \int_{\Gamma_\epsilon} \delta p^{\bot}\wedge\delta\varphi\; .
\end{align}
Equation \eqref{eq:symp_Gamma=symp_V_int} means that the
transition from the space of Cauchy data on the hyperboloid
$V_{\inter, \epsilon}$ to the space of boundary data on the cone
$\Gamma_{\epsilon}$, defined by the field dynamics, is a canonical
transformation (a {\em symplectomorphism}).

Using \eqref{eq:Lagr-ren} on the Cauchy surface $\Sigma=\{\tau =
\text{const.}\}\supset V_{\inter,\epsilon}$  we obtain:
\begin{align*}
  \varphi\Big|_{V_{\inter,\epsilon}} &=\phi\;,
  \\
    p^{\bot} = \pi^\mu\partial_\mu \rfloor \romand^4\xi \,
    \Big|_{V_{\inter,\epsilon}} &=
  \pi \romand^3\xi\;.
\end{align*}
On the other hand, on
$\Gamma_\epsilon=\{\|\xi\|=\frac{1-\epsilon}{1+\epsilon}\}$ we
have:
\begin{align*}
  \varphi\Big|_{\Gamma_{\epsilon}} &=: f\;,
  \\
    p^{\bot} = \pi^\mu\partial_\mu
    \rfloor \romand^4\xi \, \Big|_{\Gamma_{\epsilon}} &=
  \pi^k\frac{\xi_k}{\|\xi\|}\, \|\xi\|^2\romand^2\xi \romand\tau=
  (\partial_\tau \varphi) \|\xi\|^2\romand^2\xi \romand\tau =
  (\partial_\tau f) \big(\tfrac{1-\epsilon}{1+\epsilon}\big)^2
  \romand^2\xi \romand\tau\;,
\end{align*}
where $f$ is a function which lives on $\Gamma_\epsilon$.
Thus equation \eqref{eq:symp_Gamma=symp_V_int} takes the form
\begin{align}
  \notag
  \omega_{\inter,\epsilon} :=
    \int_{V_{\inter,\epsilon}} \delta \pi\wedge\delta\phi\; \romand^3\xi
    &= \int_{\Gamma_\epsilon} \partial_\tau\delta f\wedge\delta f
    \big(\tfrac{1-\epsilon}{1+\epsilon}\big)^2\romand\tau\romand^2\xi\;
    \\
    \label{eq:int_lambda>0}
    &= \int_{S^2}\int_{\lambda \geq 0} \partial_\lambda\delta f_{\epsilon}^{+}
    \wedge\delta f_{\epsilon}^{+} \romand\lambda\romand^2\sigma\; ,
\end{align}
where
\begin{align*}
    f_{\epsilon}^{+} (\lambda,\dots)
    := \frac{1-\epsilon}{1+\epsilon}\ f(\lambda+\tau,\dots)\; .
\end{align*}

Formulae \eqref{eq:int_lambda<0} and \eqref{eq:int_lambda>0} prove
that the global Cauchy data can be described by a single function
$(f_\epsilon)$, equal to $f_\epsilon^{-}$ for $\lambda <0$ and to
$f_\epsilon^{+}$ for $\lambda > 0$. Tailoring these two partial
phase spaces into a single phase space, we have to impose
compatibility condition (``corner condition") at $\lambda=0$,
namely: the symplectic form:
\begin{align}
    \omega_{\epsilon}:= \omega_{\exter,\epsilon}+\omega_{\inter,\epsilon}
    =\int_{S^2} \romand^2\sigma \int_{\lambda \in \mathbb{R}}
    \delta f_\epsilon^\prime
    \wedge\delta f_\epsilon
    \;\romand\lambda
\end{align}
must be well defined. It means that the product of function which
lives on a tube $\Real\times S^2$ and its derivative along the
tube must be integrable. Moreover, functions $f_\epsilon$ and
$\partial_\lambda f_\epsilon$ have to belong to mutually dual
spaces because they represent``positions" and ``momenta"
correspondingly. This implies that $f_\epsilon \in \Hspace^{\frac
1 2}(\Real)\otimes \Lspace^2(S^2)$ and $\partial_\lambda
f_\epsilon \in \Hspace^{-\frac 1 2}(\Real)\otimes \Lspace^2(S^2)$.

\end{document}